\documentclass{sigplanconf}
\usepackage{balance}  
\usepackage{graphics} 
\usepackage[T1]{fontenc}
\usepackage{txfonts}
\usepackage{mathptmx}
\usepackage[pdftex]{hyperref}
\usepackage{color}
\usepackage{booktabs}
\usepackage{textcomp}
\usepackage{microtype} 
\usepackage{ccicons}  

\usepackage{todonotes}

\usepackage{booktabs}
\usepackage{array}
\newcolumntype{L}[1]{>{\raggedright\let\newline\\\arraybackslash\hspace{0pt}}m{#1}}
\newcolumntype{C}[1]{>{\centering\let\newline\\\arraybackslash\hspace{0pt}}m{#1}}
\newcolumntype{R}[1]{>{\raggedleft\let\newline\\\arraybackslash\hspace{0pt}}m{#1}}

\usepackage{subcaption}
\usepackage{xspace}
\newcommand{\eg}{\emph{e.g.,}\xspace}
\newcommand{\ie}{\emph{i.e.,}\xspace}
\newcommand{\etal}{ et~al.\xspace}

 \def\plaintitle{IFTTT vs. Zapier: A Comparative Study of Trigger-Action Programming Frameworks}
\authorinfo{Amir Rahmati}
           {Stony Brook University}
           {amir.rahmati@stonybrook.edu}
\authorinfo{Earlence Fernandes}
           {University of Washington}
           {earlence@cs.washington.edu}
\authorinfo{Jaeyeon Jung}
           {Samsung}
           {jae.jung@samsung.com}
\authorinfo{Atul Prakash}
           {University of Michigan}
           {aprakash@umich.edu}           

\makeatletter
\def\url@leostyle{%
  \@ifundefined{selectfont}{
    \def\UrlFont{\sf}
  }{
    \def\UrlFont{\small\bf\ttfamily}
  }}
\makeatother
\def\pprw{8.5in}
\def\pprh{11in}

\def\draft{0} 
\def\anonymize{0}

\newcommand{\dateCollected}{September $6^{th}$, 2016\xspace}
\newcommand{\DateIftttFirstSet}{July $24^{th}$, 2016\xspace}

\newcommand{\sumTrigAct}{6406\xspace} 
\newcommand{\iftttTrig}{1239\xspace}
\newcommand{\iftttAct}{738\xspace}
\newcommand{\zapierTrig}{3884\xspace}
\newcommand{\zapierAct}{2140\xspace}

\newcommand{\sumChannels}{1051\xspace} 
\newcommand{\iftttChan}{350\xspace}
\newcommand{\zapierChan}{701\xspace}

\newcommand{\commonChannels}{47\xspace}
\newcommand{\commonTrigAct}{103\xspace}

\newcommand{\avgIFTTTTrigs}{4\xspace}
\newcommand{\avgIFTTTActs}{2\xspace}
\newcommand{\avgZapierTrigs}{6\xspace}
\newcommand{\avgZapierActs}{3\xspace}

\ifnum\draft=1
\newcommand{\Revision}{0}
\usepackage{drafthead}
\newcommand{\amir}[1]{\textbf{\color{purple}[ Amir: #1]}}
\newcommand{\ef}[1]{\textbf{\color{blue}[ EF: #1]}}
\newcommand{\ap}[1]{\textbf{\color{magenta}[ AP: #1]}}
\newcommand{\jj}[1]{\textbf{\color{magenta}[ JJ: #1]}}
\definecolor{purple}{rgb}{1,0,1}
\else
\newcommand{\amir}[1]{}
\newcommand{\ef}[1]{}
\newcommand{\ap}[1]{}
\newcommand{\jj}[1]{}
\fi

\begin{document}

\title{\plaintitle}




 \maketitle

\begin{abstract}
The growing popularity of online services and IoT platforms along with increased developer's access to devices and services through RESTful APIs is giving rise
to a new class of frameworks that support trigger-action programming. These frameworks provide an interface for end-users to bridge different RESTful APIs in a trigger-action
model and easily create automated tasks across diverse platforms. Past work has characterized the space of user-created trigger-action combinations in the context of IFTTT, a popular trigger-action framework. In this work, we characterize the space of possible functionality that such frameworks open up to end-users in the context of two major frameworks---IFTTT and Zapier---and discuss results from our comparative analysis of these frameworks. We create a snapshot of $\sumTrigAct$ triggers and actions from $\sumChannels$ channels/apps across these two frameworks and compare the available functions, distribution of channels, and functions shared between them. We examine user's ability to define their own channels, triggers, and actions; analyze the growth of these frameworks; and discuss future research opportunities in this domain.

\end{abstract}

\section{Introduction}
The migration of tools and applications to the cloud, the growth of social networks, and the recent proliferation of Internet of Things (IoT) platforms has taken our devices, services, and data online. Although these systems offer RESTful APIs that enable developers to integrate such services together to provide useful functionality (``If a door is unlocked, then turn off the oven,'' ``Alexa tell SmartThings to unlock the door,'' ``Add new posts you like on Instagram to Dropbox''), end-users have no easy way of creating their own integrations. Therefore, we are seeing an emergence of a new class of programming frameworks that offer a simple yet effective trigger-action model that enable end-users to create useful integrations on their own. Users connect triggers (\eg if door lock is engaged) and actions (\eg turn off the oven) together to create automated tasks.
Given their simplicity and effectiveness, the research community has explored various facets of these models~\cite{ur_chi14,walch2013homeblox,newman2008providing,pane2001studying}.



These frameworks allow users to achieve such automation using the \textit{channel} abstraction. A channel represents the functions of the online service on the trigger-action framework, and uses the OAuth authorization protocol to securely connect to the online service's RESTful APIs. Therefore, the channels represent the space of functionality that is possible in these programming frameworks. Prior work has explored the space of user-created trigger-action combinations (or recipes). Ur \etal analyzed the effectiveness of trigger-action programming in smart homes~\cite{ur_chi14}. Ur \etal also created a snapshot of 200,000+ recipes of IFTTT (If-This-Then-That) and characterized the space of user-created recipes~\cite{ur_chi16}. In contrast, we characterize the space of possible functionality in such frameworks by analyzing the properties of their channels. We focus on IFTTT~\cite{ifttt} and Zapier~\cite{zapier}, two popular trigger-action programming frameworks. IFTTT supports a single trigger and action combination in each task and is focused more on consumer services (\eg News, social networks) and devices (\eg Environment control, applicances) with the aim of being easy to use. Zapier~\cite{zapier} on the other hand enables creating more complex tasks and focuses its service on business needs such as project management and marketing automation.




We snapshot \sumTrigAct triggers and actions across \sumChannels channels and analyze the similarities and dissimilarities between channels in the two frameworks. We look at distribution of functions in two frameworks and examine design differences and their implications on how these systems are utilized by end-users. Our results show a trade-off between customizability and usability in these systems.

To promote future research in this domain, we are making our dataset available at
\ifnum\anonymize=0 \url{https://iotsecurity.eecs.umich.edu/} \else [Anonymized]. \fi

\section{IFTTT vs. Zapier}

\begin{figure*}[!t]
\center
\includegraphics[width=\textwidth]{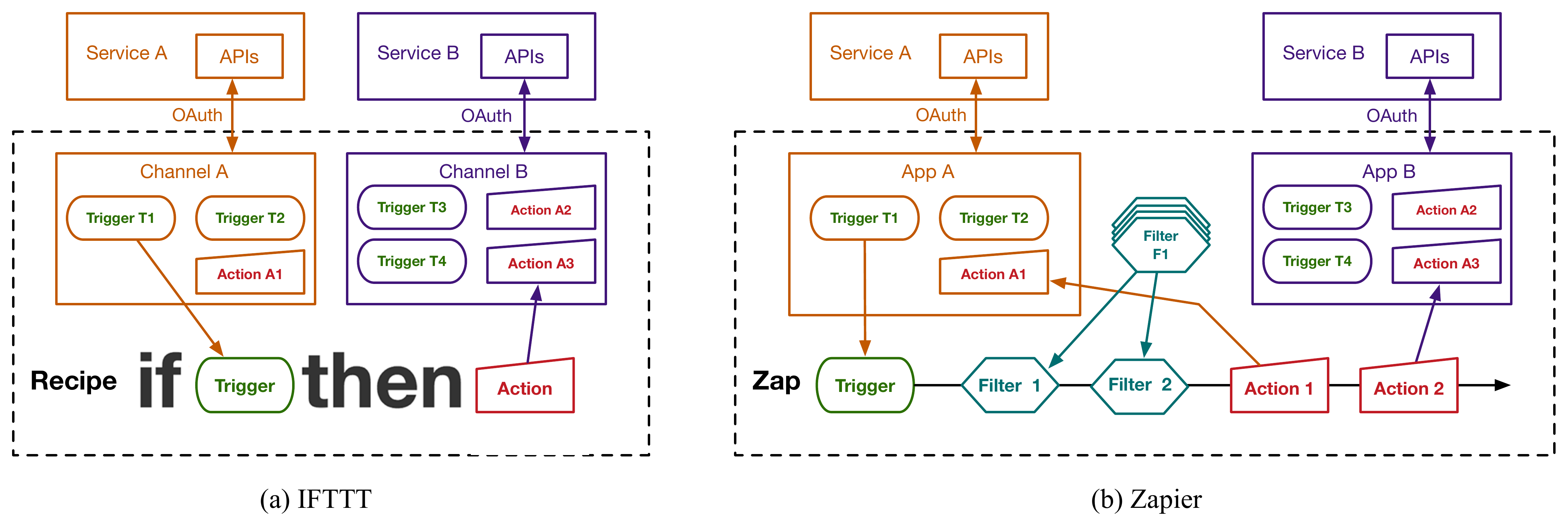}
\caption{Design overview of IFTTT and Zapier. While two designs share many characteristics (\eg definition of channels/apps and how they connect to services), they offer different levels of customizability in their definition of tasks (\ie recipes/zaps).}
\label{fig:sysFlow}
\end{figure*}

The goal of both IFTTT and Zapier frameworks is to provide an easy way for non-developers to automate activities across multiple platforms  by integrating their functionalities. Hence, these two frameworks share many design characteristics. Figure~\ref{fig:sysFlow} provides an overview of these frameworks. Both frameworks typically connect to various services and devices by first acquiring an OAuth~\cite{oauth} token. This token allows these frameworks to access the APIs provided by the service and initiate tasks without further user intervention.  
Each service is represented in these frameworks as a collection of functions. This abstraction is known as a \emph{channel} in IFTTT and an \emph{app} in Zapier. For consistency, we will call them channels from this point on.

There are two types of functions in each channel: \emph{triggers} and \emph{actions}. A trigger, is an event that occurs in the associated online service. ``A file was uploaded to a cloud drive'' or ``SmartLock was unlocked'' are examples of triggers. An action, is an operation (or set of operations) that exists in the API of the online service. Examples of actions include ``turning on or off a connected oven'' or ``sending an SMS.''

Both frameworks allow users to combine triggers and actions to define various \emph{tasks}. These tasks are known as \textit{Recipes} in IFTTT and \textit{Zaps} in Zapier. IFTTT limits each task to a single trigger and action. This provides a simplistic model that is easy to follow for the end-user. Zapier on the other hand allows multiple actions to be tied to a single trigger. This allows Zapier to support more complex operations such as \emph{searches} and modification of existing data. Zapier also provides a series of \emph{filters} that provide additional conditional (and/or) control over trigger components (\eg if ``receiving an email'' acts as a trigger, filters could include ``subject contains X'' and/or ``is from Y''). This allows for much richer customizablity but can potentially overwhelm the typical user.

There are other characteristics that make the systems different from each other. IFTTT provides a social component for its tasks where users can share their recipes with each other and it provides statistics on how many people use a recipe. These user interactions have been studied by Ur~\etal~\cite{ur_chi16}. It also provides a mobile application for both Android and iOS platforms that allow mobile services to be added as channels. Zapier on the other hand provides developers with the capability to define their own custom channels. We will describe our data collection and findings in the next section and then discuss the implications of these design differences.

\section{Methodology}
We created a snapshot of IFTTT and Zapier channels on \dateCollected. We first analyzed the network traces of the communication between the IFTTT and Zapier front-ends and their respective back-ends. This helped us identify RESTful endpoints that enabled downloading channel details. We coupled this with HTML screen scraping to create our dataset of \iftttChan IFTTT channels and \zapierChan Zapier channels. Our dataset is comprehensive---it contains information about channel name, the set of triggers and actions including human-readable descriptions, and the set of input arguments for each trigger and action.
We provide an example of channel data structure for IFTTT and Zapier along with the dataset at our website.\footnote{\ifnum\anonymize=0 \url{https://iotsecurity.eecs.umich.edu/} \else [Anonymized] \fi}

\section{Findings}
\subsection{Rapid growth}

\begin{table}[tb!]
\footnotesize
\centering
\begin{tabular}{C{2cm}|c|ccC{1cm}C{1cm}}
\toprule
\textbf{Characteristic}         & \textbf{Zapier} & \multicolumn{4}{c}{\textbf{IFTTT}} \\
\toprule
Date Collected         & \textbf{Sep'16}      & \textbf{Sep'16}    & Jul'16     &  Sep'15 \cite{ur_chi16} &    Jun'13 \cite{ur_chi14} \\ \toprule
$\#$ of Channels         &     \zapierChan  &   \iftttChan   &     293   &    220  & 69 \\
$\#$ of Trigger Channels &     696     &    285   &   238    &    177    &  52 \\
$\#$ of Action Channels  &     568    &    244     &   204     &    143  &  45 \\
$\#$ of Triggers         &     \zapierTrig  &  \iftttTrig  &     1036     &    768   &  180 \\
$\#$ of Actions          &     \zapierAct  &   \iftttAct         &    598    &    368  & 106 \\
\toprule
\end{tabular}
\caption{Numbers of channels, channels that include a trigger, channels that include an action, triggers, and actions in IFTTT and Zapier. We also include data from our snapshot of IFTTT on July, 2016 and two snapshots provided by Ur~\etal for comparison. IFTTT data shows a linear increase in number of channels, triggers and actions through time.}
\label{table:dataset}
\end{table}

Our dataset, collected on \dateCollected, includes \sumChannels channels and \sumTrigAct triggers and actions in total. 
Table~\ref{table:dataset} compares our collected dataset with a previous snapshot of IFTTT collected on \DateIftttFirstSet by our group and also IFTTT snapshots collected by Ur~\etal~\cite{ur_chi14,ur_chi16}. In terms of sheer numbers, Zapier offers more than twice the number of channels, but more than three times the number of functions. So channels in IFTTT are on average sparser than Zapier. Also based on trend data, we see a linear increase in number of channels, triggers, and actions across time. 

\subsection{Distribution of Triggers \& Actions}
Our data shows that triggers and actions are not evenly distributed among channels in either of the frameworks. Figure~\ref{fig:cdf} presents CDF of triggers and actions in IFTTT and Zapier. More than $11\%$ of triggers and channels in IFTTT are limited to the top $10$ channels and the first $70$ channels account for $51\%$ of triggers and $58\%$ of actions. Similarly in Zapier, More than $10\%$ of triggers and actions are in the top $20$ channels. There are also many channels in both systems with only a single trigger or action.
\begin{figure}[!tb]
\center
\includegraphics[width=\columnwidth]{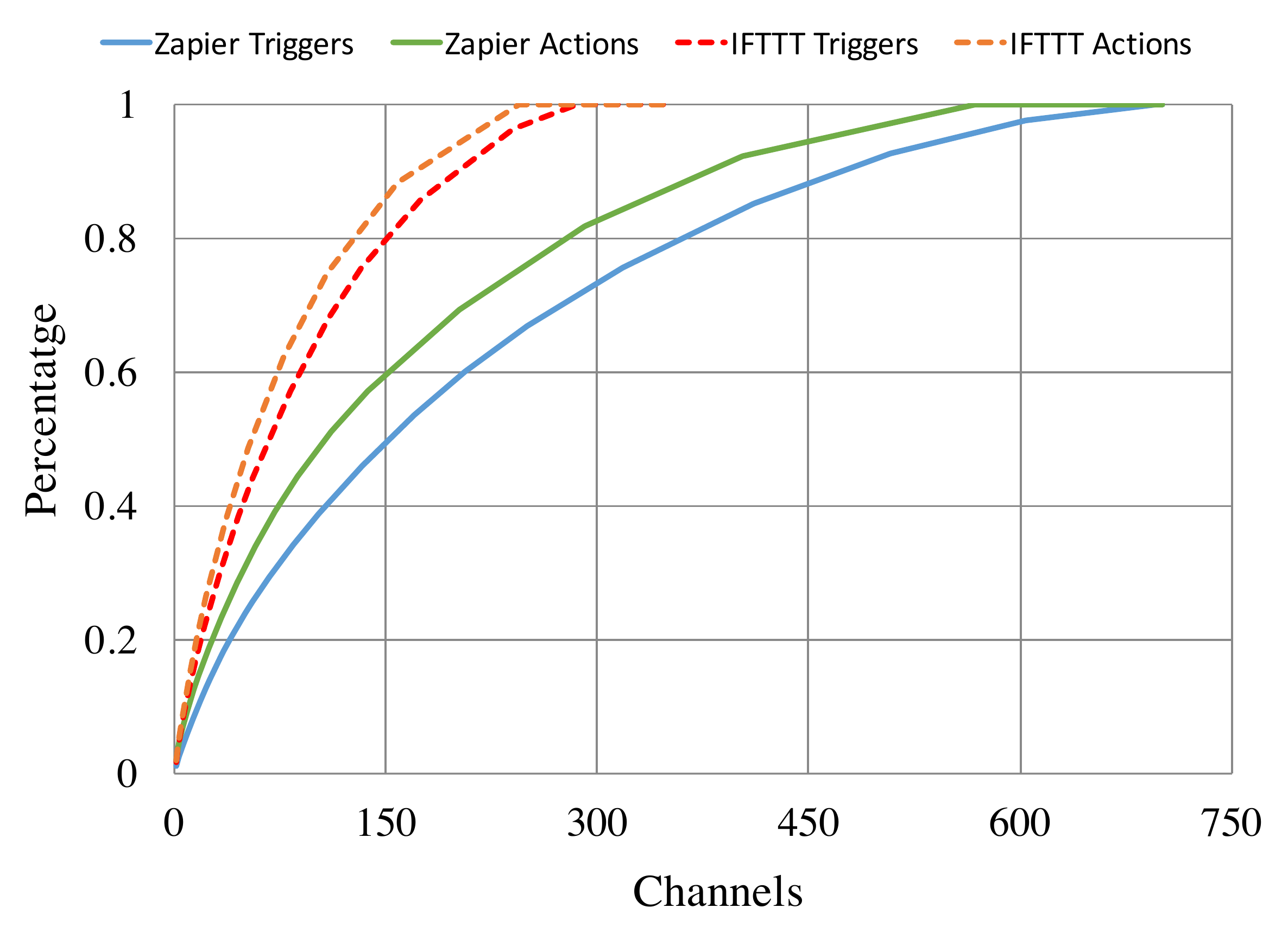}
\caption{CDF of Zapier and IFTTT triggers and actions. $50\%$ of channels in both IFTTT and Zapier hold more than $80\%$ of triggers and actions.}
\label{fig:cdf}
\end{figure}

\subsection{Difference in Target Audience}

\begin{table}[!tb]
\footnotesize
\centering

\begin{tabular}{c|C{2.5cm}c|C{2.5cm}c}
\toprule
   & \textbf{Zapier}                                 & \textbf{Count} & \textbf{IFTTT}                              & \textbf{Count} \\ \toprule
1  & Customer Relationship Management (CRM) & 44    & Environment Control and Monitoring & 39    \\ \midrule
2  & Project Management                     & 42    & Security and Monitoring Systems    & 32    \\ \midrule
3  & Marketing Automation                   & 33    & Smart Hubs and Systems             & 25    \\ \midrule
4  & Phone \& SMS                           & 28    & Appliances                         & 23    \\ \midrule
5  & Forms                                  & 24    & News and Information               & 20    \\ \midrule
6  & Email Newsletters                      & 23    & Communication                      & 12    \\ \midrule
7  & Customer Support                       & 23    & Photo and Video                    & 12    \\ \midrule
8  & Server Monitoring                      & 19    & Social Networks                    & 12    \\ \midrule
9  & Management Tools                       & 19    & Business Tools                     & 11    \\ \midrule
10 & Accounting                             & 16    & Fitness and Wearables              & 11    \\ \midrule
11 & Social Media                           & 16    & Notifications                      & 11    \\ \midrule
12 & Documents                              & 15    & Shopping                           & 11    \\ \midrule
13 & Drip Emails                            & 15    & Power Monitoring and Management    & 10    \\ \midrule
14 & Websites                               & 14    & Bookmarking                        & 8     \\ \midrule
15 & Surveys                                & 14    & Connected Car                      & 8     \\ \midrule
16 & Ecommerce                              & 14    & DIY Electronics                    & 8     \\ \midrule
17 & Developer Tools                        & 14    & Developer Tools                    & 8     \\ \midrule
18 & Scheduling                             & 14    & Mobile Devices and Accessories     & 8     \\ \midrule
19 & Payment Processing                     & 14    & Cloud Storage                      & 7     \\ \midrule
20 & Education                              & 13    & Finance and Payments               & 7  
\\ \bottomrule
\end{tabular}
\caption{Top 20 channel categories in Zapier and IFTTT. Zapier focuses more on business-oriented channels, while IFTTT focuses on consumer-oriented channels and IoT.}
\label{table:popCat}
\end{table}

Based on types of the channels available, IFTTT and Zapier serve a widely different target audience. Table~\ref{table:popCat} presents the top~20 categories and number of channels associated with them in IFTTT and Zapier. While IFTTT focuses on home-user functions such as smart home, IOT, and social networking; Zapier focuses on business functionalities such as customer and project managment, development, and marketing services. This distinction is also observable in type of triggers and actions provided by two services. Figure~\ref{fig:cloud} presents word cloud of trigger and action titles for each framework. We created the word cloud using trigger and action names from IFTTT and Zapier after filtering out common english stopwords using NLTK.\footnote{http://www.nltk.org/}
While we see words such as \textit{photo, temperature,} and  \textit{detected}, which are primarily associated with home-user, as prominent keywords in IFTTT; Zapier steers more toward keywords typical for business use. 

\begin{figure*}[t!]
    \centering
    \begin{subfigure}[t]{0.5\textwidth}
        \centering
        \includegraphics[width=\columnwidth]{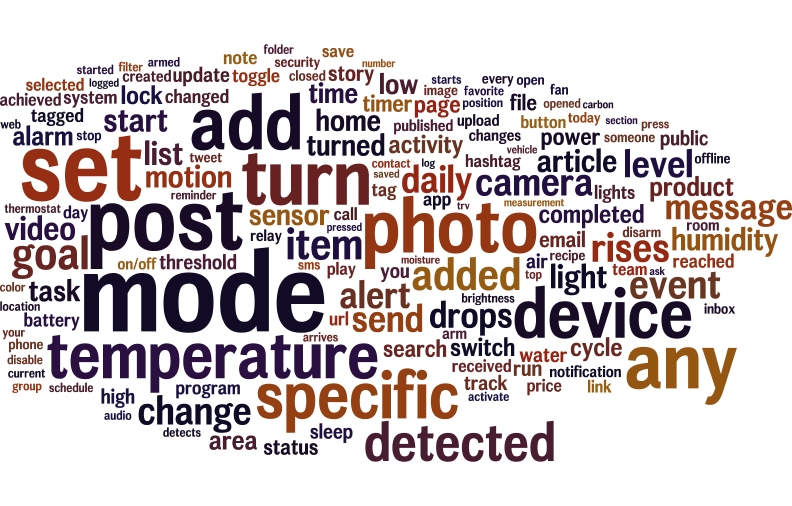}
        \caption{IFTTT}
    \end{subfigure}%
    ~ 
    \begin{subfigure}[t]{0.5\textwidth}
        \centering
        \includegraphics[width=\columnwidth]{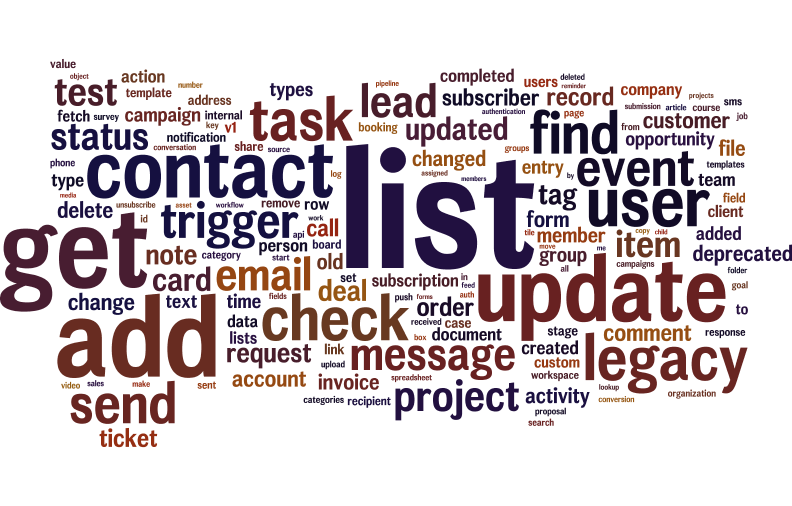}
        \caption{Zapier}
    \end{subfigure}
    \caption{Word cloud of IFTTT and Zapier trigger and action titles. Common English stopwords have been filtered from the results.}
    \label{fig:cloud}
\end{figure*}

\subsection{Shared Channels}
Even though the two frameworks seem to serve different audiences, there were still \commonChannels channels shared among them. Table~\ref{table:sharedCat} presents the top 5 categories in each framework that the shared channels fall under. These categories typically fall within domains that are shared between home and business user. We also computed lexical matches between title of triggers and actions in both services. Our results showed \commonTrigAct common functions shared in common channels between the two services. This accounts for $45\%$ and $24\%$ of functions existing for those channels in IFTTT and Zapier, respectively. 

\begin{table}[!tb]
\footnotesize
\centering
\begin{tabular}{c|C{2.5cm}c|C{2.5cm}c}
\toprule
   & \textbf{Zapier}                                 & \textbf{Count} & \textbf{IFTTT}                              & \textbf{Count} \\ \toprule
1  & Social Media & 5    & Business Tools & 7    \\ \midrule
2   & Bookmarks    & 3  & Task Management   & 6  \\ \midrule
3     & Time Tracking             & 3  & Social Networks                   & 5   \\ \midrule
4    & Devices                         & 3   & Communication                           & 4   \\ \midrule
5      & Team Chat               & 3 & Cloud Storage                                  & 3
\\ \bottomrule
\end{tabular}
\caption{Top 5 shared channel categories in Zapier and IFTTT. Common channels typically fall into categories used by both home and business users.}
\label{table:sharedCat}
\end{table}

\section{Discussion}
\label{sec:discussion}
To our knowledge, we present the first large-scale exploration of the space of functionality that trigger-action programming frameworks provide. We find that beyond targeting different application areas (\eg physical devices vs. business productivity), the two frameworks that we studied (IFTTT and Zapier) also differ in the level of customization they offer end-users during recipe creation. However, we also find that IFTTT and Zapier are capable of interoperating with each other, leading to the possibility of finding different balances between customizability, application area, and usability depending on end-user service composition scenarios.

\subsection{Customizability vs. Ease-of-Use}
Although IFTTT channels have on average $\avgIFTTTTrigs$ triggers and $\avgIFTTTActs$ actions per channel, it only supports a single trigger and action per recipe. In contrast, Zapier features $\avgZapierTrigs$ triggers and $\avgZapierActs$ actions on average per channel and supports multiple triggers and actions per zap involving AND/OR combination operators. Therefore, it would appear that IFTTT is easier to use in comparison to Zapier since the end-user has lesser cognitive load while creating trigger-action combinations. However, Ur \etal showed that often users would require multiple triggers and actions in a single recipe~\cite{ur_chi14}. Therefore, for more advanced users, IFTTT could readily support multiple triggers and actions since its channels already feature multiple operations, as our analysis shows.

\subsection{IFTTT Maker and Zapier Webhooks}
IFTTT and Zapier feature one-size-fits-all channels that can be used to make and receive HTTP web requests from arbitrary external platforms. Both IFTTT and Zapier support the standard set of HTTP verbs, while IFTTT currently supports three content-types (json, form-urlencoded, text) and Zapier supports fully custom requests. We observe that these mechanisms can be used to add additional functionality to the frameworks, although in a non-uniform way. That is, users will have their own names and semantics for the operations they enable using the Maker and Webhooks mechanisms. Finally, maker and webhooks are private to the user who created them.

\subsection{Developer-friendliness}
IFTTT currently offers an invite-only program to extend it with new channels. In contrast, Zapier offers a developer mode where channels can be created directly by users. We expect that such functionality is exercised by the advanced user. Furthermore, we note that Zapier developer mode can be used to create cyber-physical channels that are not present by default, but this would of course be relatively inaccessible to the casual user since it requires understanding RESTful APIs and authorization protocols such as OAuth.

\subsection{Open Questions}

We discuss a few open questions based on our analysis of channels in IFTTT and Zapier.

\noindent \textbf{Finding a balance between expressivity and usability:} We observed two ends of the spectrum---IFTTT offers only a single trigger and action per recipe and Zapier offers multiple. Past work has shown that multiple triggers and actions are desirable in certain settings~\cite{ur_chi14} implying that Zapier offers the better suited framework. Therefore, it could potentially be used as a tool for user studies to determine a balance between expressivity and usability.

\noindent \textbf{End-user channel development:} Although Zapier offers a developer-oriented way of adding new channels to the framework, since such frameworks are ultimately targeted towards the end-user, more research is needed to determine how to better enable end-users who are not developers to build their own channels. This would require designing point-and-click interfaces that are simple but powerful enough to achieve complex integrations that span the use of authorization protocols like OAuth and the use of RESTful APIs.



\ifnum\anonymize=0
\section{Acknowledgments}
This research is supported in part by NSF grant CNS-1318722 and by a generous gift from General Motors. Any opinions, findings, conclusions, and recommendations expressed in this work are those of the authors and do not necessarily reflect the views of the sponsors.

\fi
%
%
%
%
%

\bibliographystyle{abbrvnat}
\bibliography{sample}

\end{document}